\documentclass[aps,prl,twocolumn,showpacs,superscriptaddress,groupedaddress]{revtex4}  
\usepackage{graphicx}  % needed for figures
\usepackage{dcolumn}   % needed for some tables
\usepackage{bm}        % for math
\usepackage{amssymb}   % for math
\usepackage{color}

\begin{document}
% The following information is for internal review, please remove them for submission
% the following line is for submission, including submission to the arXiv!!
%\hspace{5.2in} \mbox{Fermilab-Pub-04/xxx-E}

\title{
Possible  Evidence for Free Precession of a Strongly Magnetized\\
Neutron Star in the Magnetar  4U 0142+61
}

\author{K. Makishima}
\affiliation{Department of Physics, Graduate School of Science,
the University of Tokyo, 7-3-1 Hongo, Bunkyo-ku, Tokyo 113-0033, Japan}
\affiliation{Research Center for the Early Universe, 
Graduate School of Science, the University of Tokyo, 
7-3-1 Hongo, Bunkyo-ku, Tokyo 113-0033, Japan}
\affiliation{MAXI team, 
RIKEN, 2-1 Hirosawa, Wako, Saitama 351-0198, Japan}
\author{T. Enoto}
\affiliation{High Energy Astrophysics Laboratory, 
RIKEN Nishina Center, 2-1 Hirosawa, Wako, Saitama 351-0198, Japan}
%RIKEN, 2-1 Hirosawa, Wako, Saitama 351-0193, Japan}
\affiliation{NASA Goddard Space Flight Center, Astrophysics Science Division, 
Code 662, Greenbelt, MD 20771, USA}
\author{J. S. Hiraga}
\affiliation{Research Center for the Early Universe, 
Graduate School of Science, the University of Tokyo, 
7-3-1 Hongo, Bunkyo-ku, Tokyo 113-0033, Japan}
\author{T. Nakano}
\affiliation{Department of Physics, Graduate School of Science,
the University of Tokyo, 7-3-1 Hongo, Bunkyo-ku, Tokyo 113-0033, Japan}
\author{K. Nakazawa}
\affiliation{Department of Physics, Graduate School of Science,
the University of Tokyo, 7-3-1 Hongo, Bunkyo-ku, Tokyo 113-0033, Japan}
\author{S. Sakurai}
\affiliation{Department of Physics, Graduate School of Science,
the University of Tokyo, 7-3-1 Hongo, Bunkyo-ku, Tokyo 113-0033, Japan}
\author{M. Sasano}
\affiliation{Department of Physics, Graduate School of Science,
the University of Tokyo, 7-3-1 Hongo, Bunkyo-ku, Tokyo 113-0033, Japan}
\author{H. Murakami}
\affiliation{Department of Physics, Graduate School of Science,
the University of Tokyo, 7-3-1 Hongo, Bunkyo-ku, Tokyo 113-0033, Japan}

\date{\today}
\begin{abstract}
 Magnetars are a special type of neutron stars,
 considered to have extreme {\it dipole} magnetic fields 
reaching $\sim 10^{11}$ T.
The magnetar 4U 0142+61, 
one of prototypes of this class,
was studied in broadband  X-rays (0.5--70 keV) 
with the  {\it Suzaku}  observatory.
In hard X-rays (15--40 keV),
its 8.69 sec pulsations  suffered
slow phase modulations by $\pm 0.7$ sec,
with a period of $\sim 15$ hours.
When this effect is interpreted as 
free precession of the neutron star,
the object  is inferred to deviate from spherical symmetry
by $\sim 1.6 \times 10^{-4}$
in its moments of inertia.
This deformation, when ascribed to magnetic pressure, 
suggests a strong {\it toroidal} magnetic field, 
$\sim10^{12}$ T, residing inside the object. 
This provides one of the first observational 
approaches towards  toroidal magnetic fields of magnetars.
%and strengthen the view that they indeed harbor
%ultra-strong internal magnetic fields.
\end{abstract}

\maketitle

%=================
%\section{Introduction}
%=================

{\em Inroduction.}---
Neutron stars (NSs) are deemed to possess strong magnetic field (MF) 
of $10^{4}-10^{11}$ T \cite{Pulsars,Chanmugam, HardingLai06}.
Their MF is attributed to, e.g.,
proton superfluids \cite{HardingLai06},
or ferromagnetism in nuclear matter  \cite{Max99, Max03},
but without clear consensus.
When studying  their magnetism,
a subclass of importance is magnetars 
\cite{Magnetar,HardingLai06,Mereghetti08},
isolated NSs believed to have
extreme dipole MFs of  
$B_{\rm d}=10^{10}-10^{11}$ T.
Their persistent and burst-like X-rays are thought
to be powered by  the MF energy,
because their luminosity much exceeds
the rate of their rotational energy loss.

We expect magnetars to harbor even stronger  {\it toroidal}  MF,
$B_{\rm t}$ \cite{Magnetar,HardingLai06,toroidal09a, toroidal09b},
because differential rotation  in their progenitors 
will tightly wind up the MF lines
during their final collapse.
We then expect some of the  internal MF lines to emerge 
from the stellar surface \cite{toroidal09b}, 
to form multipoles therein.
These expectations are supported by the recently discovered
low-$B_{\rm d}$ magnetar, SGR 0418+5729 \cite{Rea10},
because its burst activity would require MFs 
exceeding the measured $B_{\rm d}=6 \times 10^{8}$ T,
and  it shows  spectral evidence
for much stronger multipole surface MF \cite{Tiengo13}.
However, more direct estimates of $B_{\rm t}$
remained difficult.

X-ray spectra of magnetars ubiquitously consist of
a black-body-like soft  component and a distinct hard X-ray tail
\cite{Kuiper06,Enoto10c},
dominant in energies below and above $\sim 10$ keV, respectively,
both pulsed strongly at the NS's rotation period.
While the former must be thermal emission 
from two magnetic poles,
the latter may be non-thermal photons
from possibly  different regions on or around the NS \cite{Enoto10c}.
The behavior of the two components  will thus
provide clues to the magnetic structure of magnetars.

We conducted accordingly  two observations of 4U~0142+61,
one of the X-ray brightest magnetars.
It has a rotation period of 8.69 sec,
and it allowed one of the first  detections of the hard component
\cite{Kuiper06,denHartog06,denHartog08}.
On the 2nd occasion, 
its  8.69 sec pulsation in hard X-rays were
found to exhibit a slow phase modulation.
The effect may be taken as evidence for 
{\it free precession} of this NS,
and suggests its  magnetic deformation
with  $B_{\rm t} \sim 10^{12}$ T.

%===================
%\section{Observation}
%===================
%%% Para 4, 4U 0142+61
{\em Observation.}---
The two observations of 4U 0142+61 were conducted with a 2 year interval,
using the {\it Suzaku} X-ray observatory \cite{Mitsuda07}.
The soft and hard components of magnetars 
match ideally with the two {\it Suzaku} instruments;
the X-ray Imaging Spectrometer (XIS) \cite{Koyama07}
sensitive in 0.3--10 keV,
and the Hard X-ray Detector (HXD) \cite{Takahashi07}
working in 10--600 keV.
Following the first observation made in 2007 August  \cite{Paper_I},
the 2nd one reported here was performed on 2009 August 12--14,
for a gross exposure of 186 ksec (net 102 ksec).
We operated the XIS  in 1/4-frame mode 
and the HXD in normal mode,
with a time resolution of 2.0 sec and 61 $\mu$sec,
respectively.
%Our prime objective was to study 
%how the two components varied in the 2 years.

\begin{figure}[b]
  \begin{center}
   \includegraphics[width=40mm]{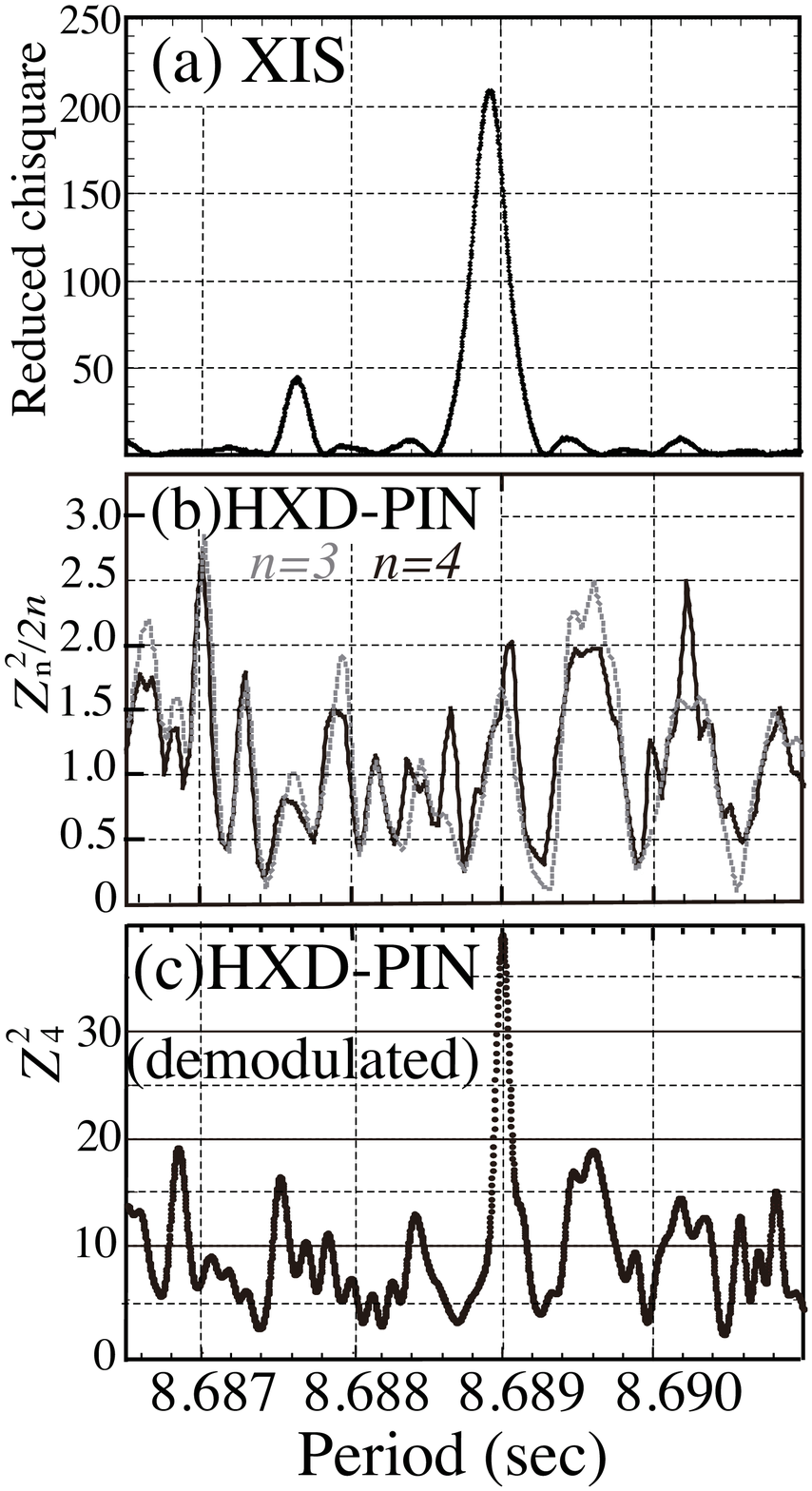}
   \hspace{1mm}
   \includegraphics[width=43mm]{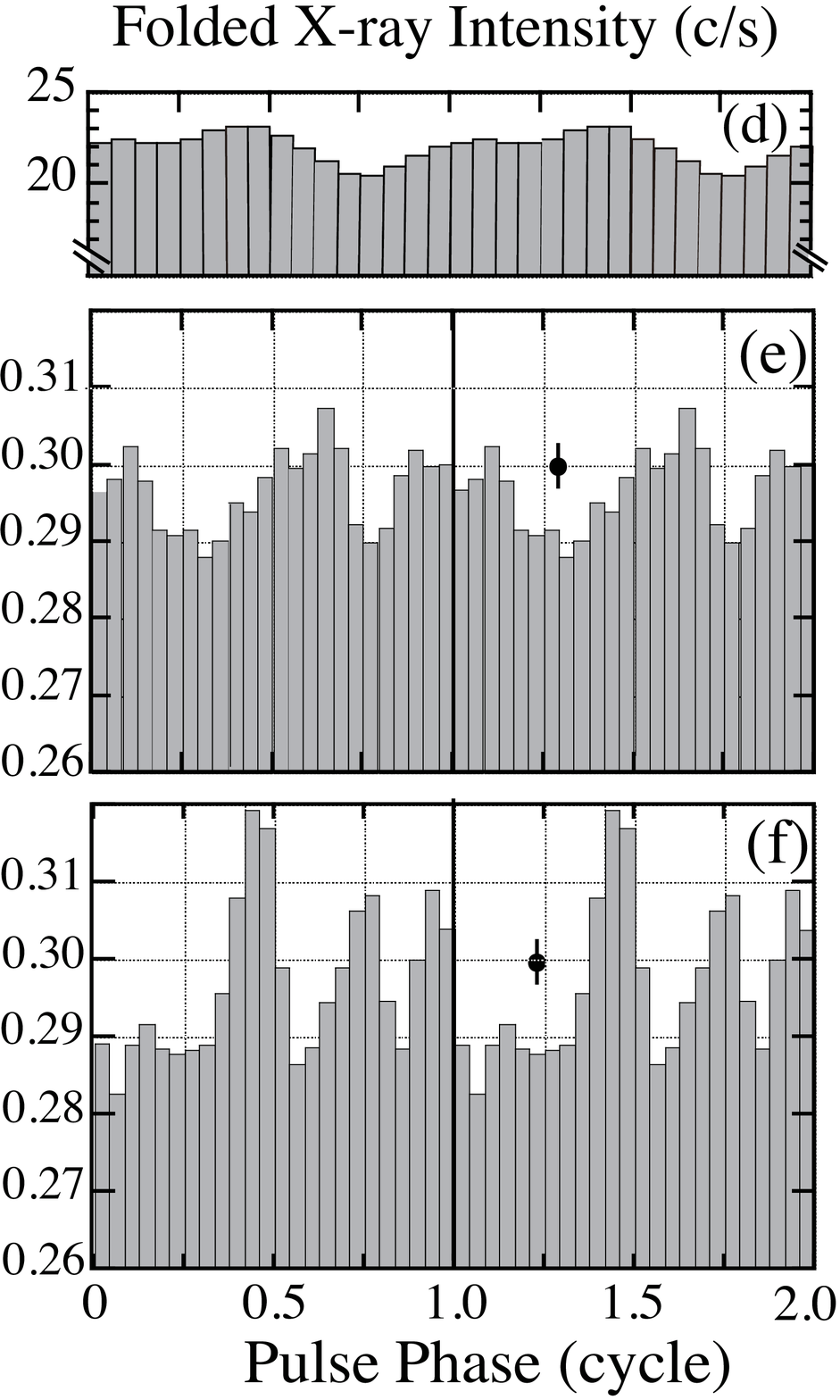}
  \caption
  {
  (a) A 1--10 keV XIS0+XIS1+XIS3 periodogram, 
  calculated via folding analysis using 16 phase bins per cycle.
  (b) Periodograms from the background-inclusive 15--40  HXD-PIN data,
  calculated using the $Z_n^2$ technique with
   $n=3$ (grey) and $n=4$ (black).
   (c) The same $Z_4^2$ periodogram as in (b),
   but after the demodulation correction employing the
   best-estimate conditions (see text).
  (d) A soft X-ray  pulse profile (two cycles), 
  obtained by folding the 1--10 keV XIS data at eq.(\ref{eq:XISpulseP}).
  (e) The  background-inclusive 15--40 keV HXD-PIN data,
  similarly folded at $P_{\rm hard}$.
  A running average over three adjacent bins was applied.
  The background  level  corresponds to 0.26 c s$^{-1}$,
  and the error bar represents statistical $\pm 1$ sigma.
%considering the running average.
  (f) The same as (e), but after the demodulation procedure.
   }
\label{fig:periodogram}
   \end{center}
\end{figure}

The source was detected  at background-removed count rates of
$6.74 \pm 0.01$ c s$^{-1}$ with the XIS in 0.4--10 keV (per camera),
and  $(3.08 \pm 0.28)\times 10^{-2}$  c s$^{-1}$
with the HXD in 15--70 keV;
both agree within $\sim 15\%$ 
with those in 2007 \cite{Paper_I}. 
Converting each photon arrival time
to that at the Solar system barycenter,
and analyzing the XIS data via epoch  folding analysis,
we detected, as shown in Fig.~\ref{fig:periodogram} (a),
the soft X-ray pulses at a barycentric period of 
\begin{equation}
P_{\rm soft} = 8.68891 \pm 0.00010 ~~{\rm sec}~.
\label{eq:XISpulseP}
\end{equation}
Together with the folded soft X-ray 
pulse profile in Fig.~\ref{fig:periodogram} (d),
this reconfirms the previous measurements  \cite{PPdot}.

% oooooooooo Figure 1 oooooooooooo
% ooooooooooooooooooooooooooooooo

%===================
%\section{Data Analysis and Results}
%===================
{\em Results.}---
We also searched the 15--40 keV HXD data 
for the expected hard X-ray pulsation \cite{Kuiper06,denHartog08,Paper_I}.
Because of lower statistics,
we employed  the $Z_n^2$ technique \cite{Zn2test94}
which
% unlike the chi-square technique,
is free from the event binning ambiguity.
If no periodicity, 
 the $Z_n^2$ values should obey 
a $\chi^2$ distribution of  $2n$ degrees of freedom (dof).
Since the hard X-ray pulse profile of 4U 0142+61 
is double-peaked \cite{Kuiper06, denHartog08}
with possible structurs  \cite{Paper_I},
we tried $n=3$ and  4.

As given in Fig.~\ref{fig:periodogram} (b),
the  HXD periodograms with $n=3$ and 4
both show a small  peak at $\sim 8.689$ sec,
at the error boundary of eq.(\ref{eq:XISpulseP}),
but its significance is rather low,
and higher peaks are seen at different periods.
%(The case with $n=2$ is similar.)
This result  was unexpected,
as the hard X-ray intensity and the observing time
were both similar to those in 2007,
wherein  the pulses were detected clearly 
both with the HXD \cite{Paper_I}  and XIS:
some changes must have taken place in the hard component.
Specifically, the hard X-ray power, 
originally at eq. (\ref{eq:XISpulseP}) (and its harmonics),
may  have been scattered out 
over a period range of Fig.~\ref{fig:periodogram},
by, e.g.,  some pulse-shape variations
as suggested before \cite{Paper_I},
or more likely, by pulse-phase modulations.
We thus came to suspect 
that the hard X-ray pulses in 2009  suffer,
for unspecified reasons,
some phase modulations.

We  assume that the  8.69 sec X-ray pulsation 
in the 2009 HXD data is  phase-modulated,
so that the peak timing $t$ of each  pulse shifts by
$\Delta t = A \sin (2\pi t/T -\phi)$,
where $T$, $A$, and $\phi$ are  the  period,
amplitude, and initial phase of the assumed modulation, respectively.
Such effects would be removed
by shifting the arrival times of individual HXD photons  by $-\Delta t$.
Employing a trial triplet $(T, A, \phi$),
we  applied these time displacements to the HXD data,
and re-calculated  the $Z_n^2$ periodograms
over an error range of eq.(\ref{eq:XISpulseP})
%(with a step $2~\mu$sec)
to see whether  the pulse significance changes.
Then, we searched for the highest pulse significance,
by scanning the three parameters over a range of 
$A=0-1.2$ sec ($0.05$ sec step), 
$\phi=0-360^\circ$ ($3^\circ-10^\circ$ step),
and $T=35-70$ ksec (1--2.5 ksec step).
%The search upper limit of $A=1.2$ was set 
%in order to limit ourselves to the cases of $A \ll P_{\rm soft}$.
%The range of $T$ was determined considering
%that the subpeaks in Fig.~\ref{fig:periodogram1}(b) 
%arise as beats between $P_{\rm soft}$ and $T$.
The $Z_n^2$ harmonic parameter was chosen to be $n=4$.
%The overall analysis is very similar to that employed by \cite{Koyama89}.

\begin{figure}[b]
  \vspace*{-3mm}
  \begin{center}
   \includegraphics[width=88mm]{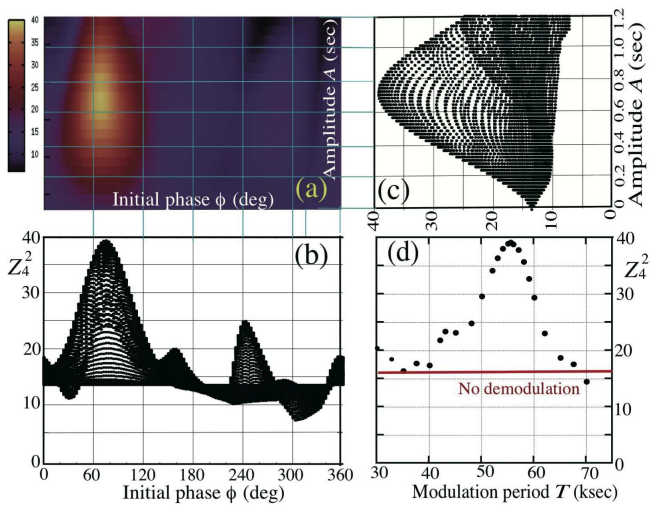}
    %% \FigureFile(width,height){filename}
  \end{center}
  \vspace*{-3mm}
  \caption{Results of the $Z_4^2$ ``demodulation" analysis,
  assuming a periodic phase shift
  in the 15--40 keV HXD pulses in 2009.
  (a) A two-dimensional  color map, on the $(\phi, A)$ plane,
   of  the $Z_4^2$  maximum found 
   over the period range of eq.(\ref{eq:XISpulseP}),
   for  $T=55.0$ ksec.
  (b) The projection of panel (a) onto the $\phi$-axis,
  where the vertical data scatter reflects differences in $A$.
  (c) The same as panel (b), but projected onto the $A$ axis.
  %Different values of $\phi$ produce the multiple traces.
  (d) The  maximum values of $Z_4^2$ found in maps as panel (a),
  plotted against  $T$.
 % The red line  indicates
  %the values of $Z_4^2$ for $A=0$ (i.e., without demodulation).
}
\label{fig:demodulation}
\end{figure}

This ``demodulation" analysis has yielded 
results  in Fig.~\ref{fig:demodulation}.
Under a condition of $T=55.0$ ksec,
the pulse significance has 
increased drastically to $Z_4^2=39.5$ (panel a)
when $\phi=75^\circ \pm 30^\circ$ (panel b)
and $A=0.7 \pm 0.3 ~{\rm sec}$ (panel c) are employed.
As in panel (d),
the modulation period was constrained
as $T=55 \pm 4~{\rm ksec}$,
where neither background variation
nor observing window has significant power.
The errors of  $\phi$,  $A$, and $T$ are represented
by the standard deviations of Gaussians
fitted to the distributions (above uniform backgrounds)
in Fig. \ref{fig:demodulation}(b)-(d).
When the data are demodulated with these conditions,
the  HXD  periodogram, Fig.~\ref{fig:periodogram} (b),
changed into Fig.~\ref{fig:periodogram} (c);
it reveals a prominent single peak at
$P_{\rm hard}  =  8.68899 (5)$ sec,
where the error was determined from the peak width in Fig.~\ref{fig:periodogram} (c).
This is  consistent with $P_{\rm soft}$ within errors.

\begin{figure}[bt]
 \begin{center}
       \includegraphics[width=45mm, height=50mm]{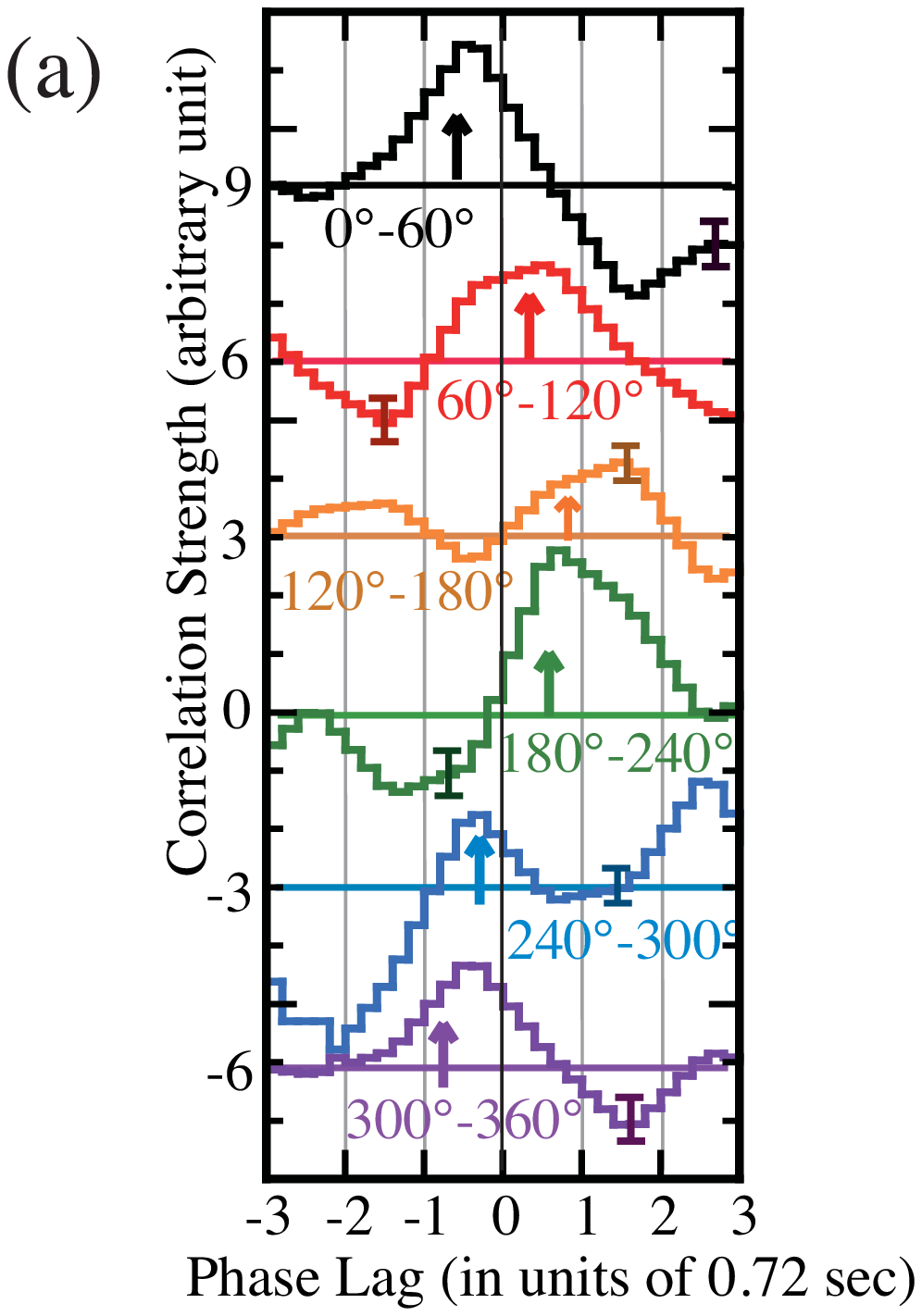}
       \vspace*{-2mm}
    \includegraphics[width=38mm]{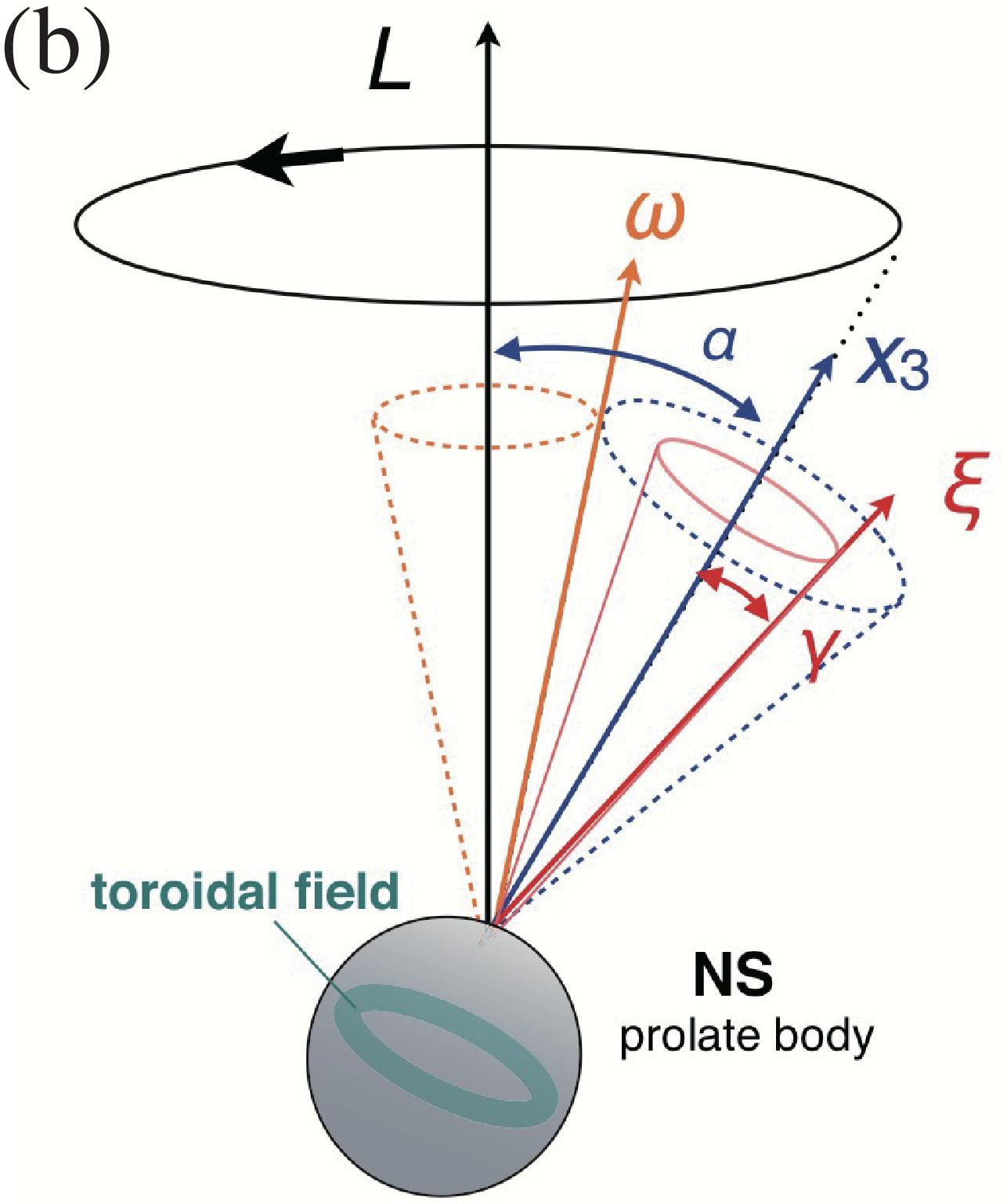}
  \end{center}
  \vspace*{-2mm}
 \caption{
 (a)  Cross correlation (interpolated and shown with offsets), 
in arbitrary unit, of the 2009 HXD profiles in Fig.\ref{fig:periodogram}(f), 
 with those accumulated (without demodulation) in 6 phases of  $T=55$ ksec.
 Arrows indicate predictions by the best demodulation parameters.
(b) An illustration of free precession of 
  an axisymmetric rigid body \cite{Butikov06,Landau}.}
  \label{fig:symmetrictop}
\end{figure}

Figures~\ref{fig:periodogram} (e) and (f), respectively,
show the HXD pulse profiles before and after the demodulation,
both folded at  $P_{\rm hard}$.
The latter exhibits a significantly larger pulse amplitude 
and richer fine structures than the former.
In addition, the HXD pulse-peak phase has been 
brought closer to  that of the XIS,
as in previous observations \cite{denHartog08}.
We further folded the 2009 HXD data (without demodulation) 
into 12 bins at $P_{\rm hard}$,  
over six separate phases of the $T=55$ ksec period,
and cross-correlated the profiles with that in Fig.~\ref{fig:periodogram}(f).
The results, given in Fig.~\ref {fig:symmetrictop}(a),
visualize the sinusoidal nature of the modulation.

% oooooooooo Figure 3 oooooooooooo

Can the peak in Fig.~\ref{fig:demodulation}  arise by chance
when considering  the many trials in $T$, $A$, $\phi$, and $P$?
As a ``control" study,
we repeated, 356 times, the same analysis as Fig.~\ref{fig:demodulation}
at periods of $P=P_{\rm hard} + \Delta P$,
scanning the offset $\Delta P$ from $-0.1$  to +0.1 sec with a  0.5 msec step
(but avoiding $\Delta P=0$  and side lobes of $P_{\rm hard}$).
Extrapolating the  obtained $Z_4^2$ grand maxima distribution
and taking its uncertainty into account,
the  chance probability to find a value of $Z_4^2 \geq 39.5$ 
in a  search like Fig.~\ref{fig:demodulation} was 
estimated as $p_{\rm z2} \equiv  (0.8-2.6) \times 10^{-3}$.
We also found that individual $Z_4^2$ values around
$P=P_{\rm hard} $ (but away from  $T=55$ ksec)
roughly obey a  $\chi^2$  distribution with 9 dof, instead of  8,
due to  the  pulsation.
We hence multiplied $p_{\rm z2} $ by a factor
$\Psi_9(39.5)/\Psi_8(39.5) = 9.4 \times 10^{-6}/4.0 \times 10^{-6} = 2.4$,
to obtain the overall chance probability of $(1.9-6.2) \times 10^{-3}$,
where $\Psi_\nu (x)$ is upper integral
for  a $\chi^2$  distribution of $\nu$ dof.
Thus, at $>99$\% confidence, 
we can exclude the case 
where the peak in Fig.~\ref{fig:demodulation}  arises via chance fluctuations.

For further examination,
we applied exactly the same demodulation search 
to three blank-sky HXD data sets,
%with exposure of 95, 10, and 41 ksec,
and another for the Crab Nebula 
%(105 ksec) 
representing high count-rate signals.
However, these data sets all gave $Z_4^2 <30$.
Since the implied  upper probability integral, 
$\Psi_8(30.0)=2.0 \times 10^{-4}$,
is still much larger than $\Psi_9(39.5)$,
the 55 ksec modulation in 4U 0142+61 is unlikely to be instrumental.
We next re-analyzed the 2007 HXD data of 4U~0142+61 with the $Z_4^2$ method,
and reconfirmed the hard X-ray pulsation with a high significance of $Z_4^2=52.0$,
at $8.68878(5)$ sec as in \cite{Paper_I}.
The 2007 data were further subjected to the  same demodulation search,
over a range of $T=55\pm 10$ ksec
which is 2.5 times wider than the 2009 uncertainty.
However, the HXD data in 2007
were not very sensitive to $A$ or $T$,
yielding a rather loose limit of $A<0.9$.
Since this limit overlaps with 
the error range of $A$  in 2009,
the 2007 HXD data accommodate the hard X-ray modulation,
but do not give an independent support to it.

Finally, the same analysis was applied to the 
two (2007 and 2009) XIS data sets of 4U~0142+61.
However,  the soft X-ray pulses on neither occasion
exhibited evidence for phase modulation 
over $T=55\pm 10$ ksec,
and  the highest pulse significance was obtained at $A^<_\sim  0.1$ sec.
Through a simulation, we  confirmed 
that this result is  not due to the insufficient 
time resolution (2.0 sec) of the XIS data: 
a 55 ksec phase modulation with $A=0.7$ sec
would have been detected 
within an error of $\Delta A \sim \pm 0.2$ sec (90\% limits).
We thus place an upper limit of  $A<0.3$ sec
for the soft X-ray pulse-phase modulation at 55 ksec.

%%%%%%%%% 4 %%%%%%%%%%%%
%\section{Discussion and Conclusions}
%%%%%%%%%%%%%%%%%%%%%
{\em Discussion.}---
The pulse-phase variation in the 2009 HXD data, 
which is rather sinusoidal (Fig. \ref{fig:symmetrictop} a),
could be due to the presence of a binary companion to the NS.
From the observed values of $T$ and $A$,
and  the canonical NS mass of $1.4~M_\odot$ 
($M_\odot$ being the Solar mass),
the putative companion is estimated to have a mass of 
$0.12 M_\odot  /\sin i$, where $i$ is the orbital inclination.
Although the implied lower-limit mass of $\sim 0.1 M_\odot$
is broadly consistent with the optical $R$-band magnitude 
of 4U~0142+61, $\sim 25$ mag \cite{optical},
the optical emission, which is  pulsed \cite{OptPulse1, OptPulse2},
is likely to emerge from a vicinity of the NS,
rather than from any companion star.
Furthermore, the absence of the same modulation in the soft X-rays
argues clearly against the binary interpretation.

% oooooooooo Figure 4 oooooooooooo
% oooooooooo Figure 4 oooooooooooo

As such, we consider that the pulse-phase modulation  
in the 2009 HXD data is more likely to reflect
intrinsic dynamics of the NS in 4U 0142+61,
especifically, free precession of an axisymmetric rigid body
which can occur without any external torque.
In  Fig. \ref{fig:symmetrictop} (b)
which illustrates  an axisymmetric NS,
let $\vec L$ be its angular momentum vector fixed to the inertial frame,
and  $\hat{x}_3$  a unit vector describing its axis of symmetry
which we may identify with the dipolar magnetic axis.
The star's asphericity
is expressed by a quantity $\epsilon \equiv (I_1 - I_3)/I_3$,
where $I_3$ is the moment of inertia  around  $\hat{x}_3$,
and $I_1$  that around axes orthogonal to $\hat{x}_3$.
If $\epsilon  \ne 0$,
the  $\hat{x}_3$ axis rotates around $\vec L$
at  a constant period $P_1=2 \pi I_1/L$,
with a constant ``wobbling" angle $\alpha$ 
to $\vec L$ \cite{Landau, Butikov06, Maggiore07}.
Likewise, the  instantaneous rotation vector $\vec{\omega}$, 
co-planar with $\vec{L}$ and $\hat{x}_3$,
rotates around $\vec{L}$,
keeping a constant %length $\omega=L/I_1$, and a 
angle  $ \approx \epsilon \sin\alpha$ to $\vec L$.

Suppose that the NS emits  photons through which we observe it,
and express the direction of maximum photon emissivity 
by a unit vector  $\hat{\xi}$  fixed to the NS.
When $\hat{\xi}=\hat{x}_3$,
the photons will  reach us in
periodic pulses with the period $P_1$.
However, if $\hat{\xi}$ is tilted from $\hat{x}_3$ by a finite angle $\gamma$,
then $\hat{\xi}$ will slowly rotate around $\hat{x}_3$
relative to the $\vec L$-$\vec\omega$-$\hat{x}_3$ plane,
with a ``slip period" $Q=P_1/\epsilon =[(1/P_3  -1/P_1)]^{-1}$
where  $P_3 \equiv 2\pi I_3/L$ \cite{Landau,Butikov06,Maggiore07}.
The observed pulse arrival times then
become subject to some jitter \cite{Butikov06},
and the effect can be approximated as 
a sinusoidal phase  modulation in the regular pulsation,
just as seen in Fig. \ref{fig:symmetrictop}(a).

The results from the 2009 HXD data  thus allow
an interpretation in terms of free precession of the NS with $T=Q$.
The necessary condition of $\gamma \neq 0$ can be fulfilled 
if , e.g., the MF has multipole  components \cite{Magnetar, Tiengo13} (\S 1).
Due to some re-arrangement of the magnetic configuration,
the hard X-ray emitting regions may wander around on the star, 
causing both $\gamma \sim 2\pi a/P_1 \sin(\alpha)$ 
and $\phi$ to change with time.
The  behavior of the two HXD datasets can be explained 
if  the hard X-ray beam pattern was sharper with $\gamma \ne 0$ in 2009, 
while  broader in 2007 with a smaller value of $\gamma$. 
The absence of the same modulation in the 
two XIS data sets can be explained 
if the soft X-ray emission comes form regions 
more symmetric ($\gamma \sim 0$) around $\hat{x}_3$,
and/or in a broader beam.
However, other senarios remain;
e.g., the putative motion of the hard X-ray source
itself could  produce red noise in the pulse phase,
which mimics the 55 ksec periodicity.

If we employ the precession interpretation,
the best-fit demodulation parameters yield
$|\epsilon|= P_1/Q = P_{\rm hard}/T = 1.6 \times 10^{-4}$.
Although  we cannot tell
whether the object is prolate ($\epsilon > 0$) or oblate ($\epsilon < 0$),
the former is more likely,
because internal dissipation will increase  $\alpha$ if  $\epsilon > 0$,
while decrease if $\epsilon < 0$.
The strong X-ray pulses observed from nearly  all magnetars, 
implying  $\alpha \ne 0$, suggests  $\epsilon > 0$.
Since  an NS with $B_{\rm t}  \gg B_{\rm d}$ 
would be deformed into a prolate shape with
$\epsilon  \sim +1.0\times 10^{-4} (B_{\rm t}/10^{12}{\rm T})^2$
\cite{Cutler02,Gualtieri11,precession_interior},
the derived estimate of $\epsilon \sim 1 \times 10^{-4}$ can be explained
by a toroidal magnetic field of $B_{\rm t}\sim 1 \times 10^{12}$ T.
It  much exceeds the value of
$B_{\rm d} = 1.3 \times 10^{10}$ T
derived from the pulse period and period derivative  of this NS \cite{PPdot}.

So far, there have been reports of possible detections
of precession from accretion-powered pulsars \cite {Precession09},
and fast-rotating pulsars \cite{Precession88, Precession01,Precession03,Precession06}.
However, the effects in the former objects 
should be  regarded as  {\it forced} precession,
considering strong torque from the accreting matter.
Similarly, the latter objects would not easily show free precession,
since they must be deformed into oblate shapes under centrifugal force,
and hence the precession would soon be damped  \cite{Cutler02},
making the reports somewhat unconvincing \cite{Ruderman06}.
In contrast, 
the present detection is considered more promising
from the stability argument made above.
(Decay in $\alpha$ due to emission of gravitational waves
is expected to be insignificant \cite{Maggiore07}.)

Supposing that our interpretation 
correctly account for the observation,
several astrophysical implications  follow.
(1) It provides one of the first observational 
clues to $B_{\rm t}$ inside NSs,
which is much more difficult to estimate
than $B_{\rm d}$ \cite{Pulsars,Max99}.
(2) The relatively large value of $\epsilon$
supports the view
that magnetars have
 $|B_{\rm t}|>|B_{\rm d}|$  \cite{Cutler02,Gualtieri11}.
(3) The differences  between the soft and hard X-ray 
components suggests their distinct emission regions \cite{Enoto10c}.
(4) Further studies of this object,
and other similar ones,
will provide valuable information on the NS interior
\cite{Cutler02,precession_interior,Lattimer07},
and  prospects for  gravitational-wave emission
\cite{Maggiore07,Gualtieri11,Abadie11}.
%Kashiyama+Iona11}.

In summary, we suggest that;
the NS in 4U 0142+61 is  deformed 
with $|\epsilon|=  1.6 \times 10^{-4}$;
the hard X-ray emission region moves, to some extent, on/around this NS;
and  the  NS harbors an intense toroidal field of $B_{\rm t} \sim 10^{12}$ T.

 % ooooooooooooooooooooooooooooooo

\end{document}